\documentclass[11pt]{article}
\pdfoutput=1 
\usepackage{jheppub} 
\usepackage[T1]{fontenc} 
\usepackage{slashed} 
\usepackage[utf8]{inputenc} 

\newcommand{\gev}{\text{GeV}}
\newcommand{\ifb}{\text{fb}^{-1}}

\title{Minimal Models of Loop-Induced Higgs Lepton Flavor Violation}

\author{Carlos Alvarado,}
\author{Rodolfo M. Capdevilla,}
\author{Antonio Delgado and}
\author{Adam Martin}

\affiliation{Department of Physics, University of Notre Dame\\225 Nieuwland Science Hall, Notre Dame, IN 46556, U.S.A.}

\emailAdd{calvara1@nd.edu}
\emailAdd{rcapdevi@nd.edu}
\emailAdd{adelgad2@nd.edu}
\emailAdd{amarti41@nd.edu}

\abstract{The LHC has recently reported a slight excess in the $h\rightarrow \tau \mu$ channel. If this lepton flavor violating (LFV) decay is confirmed, an extension of the Standard Model (SM) will be required to explain it. In this paper we investigate two different possibilities to accommodate such a LFV process: the first scenario is based on flavor off-diagonal $A$-terms in the Minimal Supersymmetric Standard Model (MSSM), and the second is a model where the Higgs couples to new vectorlike fermions that couple to the SM leptons through a LFV four fermion interaction. In the supersymmetric model, we find that the sizes of the $A$-terms needed to accommodate the $h\rightarrow \tau\mu$ excess are in conflict with charge- and color-breaking vacuum constraints. In the second model, the excess can be successfully explained while satisfying all other flavor constrains, with order one couplings, vectorlike fermion masses as low as 15 TeV, and a UV scale higher than 35 TeV.}

\begin{document} 
\maketitle
\flushbottom

\section{Introduction}   
\label{sec:Intro}

The LHC has discovered a $125\, \gev$ scalar particle with properties consistent with the Standard Model (SM) Higgs boson. Post discovery, ATLAS and CMS efforts have shifted to detailed measurement of the Higgs couplings to fermions and gauge bosons~\cite{Khachatryan:2014jba,Aad:2014xva,Khachatryan:2014aep}, and this effort will continue into Run II. As of the end of Run I, with near $20\, \ifb$ at 8 TeV center of mass energy, both ATLAS and CMS have reported an excess of events in $ p p \rightarrow \tau \mu$~\cite{Aad:2015gha,Khachatryan:2015kon}; the net significance of the excess is $2.6~\sigma$, broken down into $1.3~\sigma$ for ATLAS and $2.4~\sigma$ for CMS. If the events are interpreted as coming from a lepton flavor violating (LFV) Higgs decay $h \rightarrow \tau \mu$, the excess is best fit by the branching fraction 
\begin{equation}
\mathcal{B}r(h\rightarrow \tau \mu)=0.82_{-0.32}^{+0.33}\%
\label{BFcombined}
\end{equation}

Although lepton flavor violation within the SM is firmly established in light of neutrino oscillations and is incorporated with the PMNS matrix, the SM contributions to LFV Higgs decays are proportional to the neutrino masses and are thus completely negligible. Therefore, if the $h\rightarrow \tau \mu$ signal is confirmed by more data in Run II, it would represent the first evidence of physics beyond the SM at the electroweak scale.

While more data is needed to reveal the veracity of the excess, the present result is a venue for new physics model building -- already explored in the context of extended Higgs sectors (2HDM)~\cite{Sierra:2014nqa,Altmannshofer:2015esa,Harnik:2012pb,Dodge2015,Kobayashi,Mao2015,Crivellin2015,Liu,Botella,Rosado2015,Ko,Bertuzzo,Dorsner:2015mja,Han:2016bvl,Das:2015zwa,DiazCruz:1999xe}, dimension six operators~\cite{Goudelis:2011un,Bhattacherjee,He,Cai2006,Pruna,Crivellin:2013hpa,Crivellin:2014cta}, leptonic extensions of the Standard Model~\cite{Baek,Cheung2015,Bizot,Pilaftsis:1992st,Korner:1992zk,Arganda:2004bz}, composite Higgs models~\cite{Altmannshofer:2015esa,Harnik:2012pb,Falkowski:2013jya}, discrete flavor symmetries~\cite{Heeck:2014qea,Varzielas:2015joa}, some exotic scenarios with extra dimensions~\cite{Harnik:2012pb}, axions~\cite{Chiang:2015cba}, the inverse seesaw model~\cite{Arganda:2014dta}, lepton-flavored dark matter~\cite{Baek:2015fma}, and supersymmetry~\cite{Arana-Catania:2013xma,Arganda:2015uca,Aloni:2015wvn,DiazCruz:2002er,Arganda:2004bz} (including $R$-parity violation~\cite{Vicente:2015cka} or an inverse seesaw mechanism~\cite{Arganda:2015naa}).

In this paper we study two SM extensions, one supersymmetric and one non-supersymmetric. In both cases, we look for regions of the parameter space where the rate $h\rightarrow \tau \mu$ as given by Eq.~(\ref{BFcombined}) can be accommodated while simultaneously respecting related bounds on the lepton flavor violating processes $\tau \to \mu \gamma$. 

In the supersymmetric model with $A$-term-driven LFV, we find it is possible to reach the best fit for $h\to \tau \mu$ only by considering small $\tan \beta$ and large values (yet perturbatively safe) of the $A$-term-to-slepton soft mass ratio. However, this model is affected by stability issues, i.e. the bounds are satisfied in a region of the parameter space where charge- and color-breaking minima develop~\cite{Casas:1996de,Drees:1985ie}.

In the non-supersymmetric model, the effective LFV Higgs interactions are induced via loops of new vectorlike fermions. This loop origin of the LFV terms is distinct from other models of vectorlike fermions, which rely on direct (i.e. dimension $\le\, 4$) couplings between SM leptons and the vectorlike matter~\cite{Falkowski:2013jya}. In addition to studying the compatibility of $h \rightarrow \tau\mu$ and $\tau \rightarrow \mu \gamma$, we analyze other LFV effects involving muons. In particular, we analyze how the decay $\mu \rightarrow e\gamma$, the most stringently bounded LFV process, fits into our framework. We find that the vectorlike fermion model can be consistent with all radiative LFV bounds, provided we make an additional assumption on the ratio of the $\tau-\mu$ to $\mu-e$ four-fermion couplings. We also check the limits on $\mu \rightarrow eee$ and $\mu-e$ conversion in nuclei and find them to be less constraining than the LFV radiative decay bounds. Putting these observations together, we find that this vectorlike fermion scenario is able to accommodate the $h \to \tau \mu$ excess in a way consistent with low-energy LFV constraints. In contrast to models such as Ref.~\cite{Falkowski:2013jya}, where vectorlike matter lies in the $\mathcal{O}(100\text{ GeV})$ ballpark, the heavy fermions in our scenario must be much heavier than the electroweak scale, tens of TeV, making direct production impossible at the LHC. 

The rest of this paper is organized as follows: in Sec.~\ref{sec:Models} we formulate each of the models and describe the stability issues of the supersymmetric one. Next (Sec.~\ref{sec:Numerical}), we go into the details of the non-supersymmetric, four-fermion interaction model. We present the set of benchmark parameters, and show the parameter space where the required $h\rightarrow \tau \mu$ rate can be obtained. For the same parameter set, we give numerical estimates of the $\tau \rightarrow \mu \gamma$ and $\mu \rightarrow e\gamma$ rates  and show where they are consistent with current constraints and expected constraints from future experiments. We also briefly comment on how the bounds change if some of the LFV couplings are only generated radiatively. Finally, we give our conclusions in Sec.~\ref{sec:Conclusion}. The Appendix \ref{sec:AppendixFunctions} displays the explicit loop functions used through the paper, and some comments on the form of the effective Lagrangian and hypercharge choices are presented in Appendix \ref{sec:AppendixChoice}.

\section{The models}   
\label{sec:Models} 

\subsection{The $A$-term-driven model}
\label{subsec:AtermModel}

In this section, we attempt to explain the excess in Eq.~(\ref{BFcombined}) using a flavor violating MSSM setup. The MSSM includes general flavor structures in the soft breaking Lagrangian and in the Yukawas. Since off-diagonal Yukawa couplings usually induce unacceptably large contributions to flavor changing neutral currents~\cite{Sher:1998sj, Cvetic:1998uw, PhysRevD.15.1958}, we will focus on flavor violation from the soft terms. These terms include the slepton mixing matrices $(m_{\widetilde{L}}^{2})_{ij}$, $(m_{\widetilde{E}}^{2})_{ij}$ and slepton $A$-terms, $(A_{\ell})_{ij}$; flavor violation is encoded in the off-diagonal entries of these matrices. Flavor structure in the supersymmetry breaking parameters induces LFV Higgs decays at loop level through triangle diagrams involving two slepton propagators and a single Higgsino/gaugino propagator; see Fig.~\ref{Fig:AtermDiagram}. To connect the loop to a Higgs requires a tri-scalar Higgs-slepton-slepton vertex. Within the MSSM there are tri-scalar vertices in the superpotential, proportional to $\mu\, \sin\beta\, y_{\ell}$, and in the supersymmetry breaking sector, proportional to $A_{\ell}\, \cos{\beta}$; here, $\mu$ is the Higgsino mass parameter and the $\beta$ dependence is set by which Higgs doublet is involved in the vertex. Notice that the superpotential tri-scalar interaction is suppressed by the lepton Yukawa coupling, and therefore the LFV diagrams proceeding through it, such as when the flavor violation resides in the slepton soft masses alone, will be suppressed as well. Diagrams with flavor violation directly in the $A$-term tri-scalar interaction do not have this Yukawa suppression.

Several recent works have aimed at reproducing the $h \to \tau \mu$ result (Eq.~(\ref{BFcombined})) in different corners of the MSSM. The authors of Refs.~\cite{Arana-Catania:2013xma,Arganda:2015uca} studied the contributions from both flavor off-diagonal slepton soft mass parameters and $A$-terms, concluding that the contributions from $A$-terms are more significant than those from LFV slepton masses in the low $\tan\beta$ regime. These works, however, do not report values of the parameter space for which $\mathcal{B}r(h\rightarrow \tau\mu)$ lies within the best fit branching fraction. On the other hand, the analysis in Ref.~\cite{Aloni:2015wvn} found that $A$-term-driven LFV cannot reach the best fit value for large $\tan\beta$ and with $A$-terms saturating the perturbativity bound $A_{ij}s_{\alpha}\lesssim 4\pi m_{\widetilde{l}}$. Instead, their best fit is achieved when LFV comes from slepton soft masses with large $\tan\beta$, but at the cost of employing an extremely large $\mu$ parameter $\mathcal{O}(100$ TeV).

As the diagrams with $A$-terms are the ones that contribute the most to the LFV amplitudes for values of $\mu$ of order of the EW scale, in this work we will focus exclusively on $A$-term-driven slepton LFV effects and assume strictly diagonal Yukawa matrices. As our interest in electroweak scale supersymmetry is purely to generate LFV processes, we will assume a simplified spectrum where the bino and sleptons are the only light superpartners and the other gauginos/higgsinos, Higgs bosons, and squarks are decoupled. In practice, this means that  $A$-term LFV amplitude(s) containing loops of electroweakinos other than the bino can be neglected. For $h \to \tau \mu$, the $A$-term-driven amplitude mediated by a bino is shown below in Fig.~\ref{Fig:AtermDiagram}; to close the loop, the bino is connected to a mu-slepton and tau-slepton, and the Higgs field involved is exclusively $H_{d}$.
\begin{figure}[h!]
\centering
\includegraphics[scale=0.5]{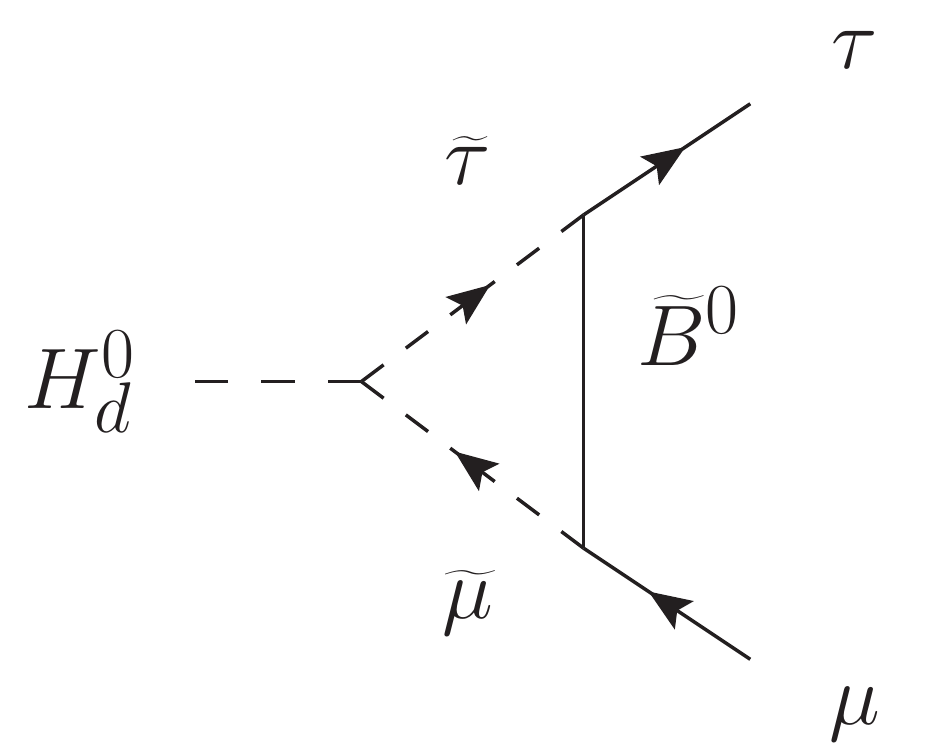}
\caption{Lepton-violating Higgs decay driven by $A$-term.}
\label{Fig:AtermDiagram}
\end{figure}
Strictly speaking, there are other diagrams coming from off-diagonal field renormalization of the lepton legs \cite{Aloni:2015wvn,Arganda:2015uca}. Yet, we stick just to the $A$-term triangle diagram because the extra diagrams will further reduce the value of the LFV branching ratio\footnote{We are thankful to  Emmanuel Stamou and Ernesto Arganda for bringing this point to our attention.}. Therefore our overestimation with the triangle diagram only will be enough to show  later the conflict between LFV $A$-terms and the charge-breaking bounds of the scalar potential.

We compute this amplitude in the mass-insertion approximation, valid provided $(A_{l})_{ij}$ $v\cos{\beta}$ $\ll$ $m_{\widetilde{l}}^{2}$. A full analysis of the LFV decays of the Higgs sector in the mass-insertion approximation and a comparison with the exact results can be found in Ref.~\cite{Arganda:2015uca}. To increase the rate, we take $(m_{\widetilde{L}})_{22}=(m_{\widetilde{E}})_{33}\equiv m_{\widetilde{l}}$. The amplitude for $h\rightarrow \tau \mu$ (not distinguishing between $\mu^{-}\tau^{+}$ and $\mu^{+}\tau^{-}$) in our simplified setup is:
\begin{equation}
\Gamma(h\rightarrow \tau \mu)=2\dfrac{m_{h}}{16\pi}\left( \dfrac{1}{16\pi^{2}} \right)^{2}\left[ \dfrac{4\pi \alpha}{c_{W}^{2}} \cos{\beta} \right]^{2}w^{2}H^{2}(r)r^{2},\label{WidthAtermhTauMu}
\end{equation}
where $\alpha$ and $c_{W}$ are the fine structure constant and the cosine of the electroweak angle, $w=|A_{\tau \mu}|/m_{\widetilde{l}}$~, and $H(r)$ is a loop function which depends on the dimensionless bino-slepton mass ratio $r\equiv m_{\widetilde{l}}/m_{\widetilde{B}}$ and is $\mathcal{O}(1)$ for the spectra we are working with (see Appendix \ref{sec:AppendixFunctions} for the explicit form of $H(r)$). Note that the branching ratio obtained from (\ref{WidthAtermhTauMu}) depends solely on the mass ratios $r$ and $w$, so bounds on $\Gamma(h\rightarrow \tau \mu)$ do not point a specific mass scale. As such, fixing $r$ and $w$ to satisfy Eq.~(\ref{BFcombined}) leaves us with the freedom to dial one of the supersymmetry breaking mass parameters ($|A_{\tau \mu}|,m_{\widetilde{l}}$ or $m_{\widetilde{B}}$) to accommodate $\tau \rightarrow \mu \gamma$. Specifically, $r=0.47$ helps maximizing the loop function $H(r)$.

In addition to flavor, another constraint we must be mindful of is vacuum stability, especially in models with multiple interacting scalars. Specifically, in the context of the MSSM, it is well known that large values of the tri-scalar $A$-terms can cause charge- or color-breaking minima to develop\footnote{Large flavor-diagonal $A$-terms would also be a problem for they can add large contributions to $h\rightarrow l_il_i$ beyond the experimental constraints.}. For flavor-violating $A$-terms, analytical bounds have been derived in Refs.~\cite{Casas:1996de,Drees:1985ie},
\begin{equation}
\left| A_{23}^{(e)} \right|\leq y_{\tau}\sqrt{m_{H_{d}}^{2}+m_{\widetilde{L}_{2}}^{2}+m_{\widetilde{E}_{3}}^{2}}~,
\label{boundA23}
\end{equation}
where $y_{\tau}$ is the MSSM tau Yukawa. It must be pointed out that bounds of the kind of (\ref{boundA23}) and related ones are conservative, and an analysis of the full theory and SUSY breaking details are needed in order to determine the true global minimum and its stability. These considerations are beyond the scope of the present work.

For equal slepton soft mass parameters, Eq.~(\ref{boundA23}) implies $|A_{23}^{(e)}|/m_{\widetilde{l}}\leq y_{\tau}\sqrt{2+m_{H_{d}}^2/m_{\widetilde{l}}^2}$. On the other hand, in the rate (\ref{WidthAtermhTauMu}) $A_{23}^{(e)}/m_{\widetilde{l}} \approx 12$, $\tan\beta=2$ are required in order to reach the best fit for $h\rightarrow \tau\mu$. In order for such large $A^{(e)}_{23}$ to be consistent with Eq.~(\ref{boundA23}), we need $m_{H_{d}}^2\geq 10^6\,m_{\widetilde{l}}^2~$ -- a huge separation among soft masses. While mathematically possible, such disparate soft masses push us outside the range of applicability of Eq.~(\ref{boundA23}); a $H_d$ field so much heavier than the other superpartners should have been integrated out and and the stability bounds applied to the resulting effective theory. Deviating from equal slepton masses only worsens the situation, as $\Gamma(h \to \tau \mu)$ decreases for $(m_{\widetilde{L}})_{22} \ne (m_{\widetilde{E}})_{33}$. Thus, we conclude that there is a clear tension between the size of the tri-scalar coupling needed to accommodate the $h\rightarrow \tau \mu$ excess and the values allowed by charge-breaking minima bounds in setups with $A$-term-driven LFV. We point out that this situation is not specific to the MSSM, but is a generic issue in models with tri-scalar interactions and calls for a careful search of true stable neutral vacua. 
Finally, while the disparity between the $w$ values that reproduce Eq.~(\ref{BFcombined}) and values satisfying inequality Eq.~(\ref{boundA23}) means the current $h \rightarrow \tau \mu$ excess cannot be reproduced in the flavor violating $A$-term MSSM, this does not exclude the possibility of restoring acceptable neutral vacua through extensions of the MSSM that involve additional interactions and/or superfields. Describing such extensions is beyond the scope of this work and we will not consider flavor violating $A$-term models -- or any other supersymmetric model -- further. 

\subsection{The four-fermion interaction model}
\label{subsec:FermionmModel}

We now present a setup in which LFV is mediated by vectorlike fermionic states. Vectorlike fermions have been incorporated in prior studies of LFV Higgs physics, especially in the context of compositeness \cite{delAguila:2011wk,Niehoff:2015bfa,Martin:2009bg}. We will not rely on tree-level Yukawa couplings between the extra matter and the SM leptons to generate Eq.~(\ref{BFcombined}). Instead, we will generate an effective LFV Yukawa vertex at loop level using higher-dimensional operators.

Let us write down the Lagrangian (in two-component spinor notation) responsible for these LFV effective interactions\footnote{In equations (\ref{LagrangianYukawa}) and (\ref{Lagrangian4f}), proper contractions between $SU(2)_L$ doublets are implicit.}:
\begin{equation}
\mathcal{L}_\text{Yuk}
= \dfrac{y}{\sqrt{2}}H(\psi^{c} \chi)+\dfrac{y'}{\sqrt{2}}H^{\dag}(\psi \chi^{c})+\text{h.c.}
\label{LagrangianYukawa}
\end{equation}
\begin{equation}
\mathcal{L}_\text{4f}
= \dfrac{1}{\Lambda^{2}}\biggl[ (\lambda_{1})_{ij}\epsilon^{\alpha \beta}\epsilon^{\rho \sigma}+(\lambda_{2})_{ij}\epsilon^{\alpha \rho}\epsilon^{\beta \sigma}]\psi_{\alpha}^{c} L_{\beta}^{i} \chi_{\rho} e_{\sigma}^{cj}+(\lambda_{3})_{ij}
(\psi \chi^{c})(L^{i\dag} e^{cj\dag})+\text{h.c.} \biggr],
\label{Lagrangian4f}
\end{equation}
where Greek indices are Lorentz indices, $i$ and $j$ are flavor indices, and $L\, (e^c)$ are the SM lepton $SU(2)_L$ doublets (singlets). The structure of $\mathcal{L}_{4\text{f}}$ is a result of the charge assignments and the correct counting of all the linearly independent contractions between the different fermion~\cite{Lehman:2015via}. Following the effective field theory approach, the operators in Eq.~(\ref{LagrangianYukawa}) should be thought of as independent, so then there is no reason for the Yukawa couplings or $\lambda_{i}$ couplings to be identical.

By first coupling the Higgs to vectorlike states $\psi,\chi$ with a Yukawa vertex, and then closing a loop by coupling these new fermions to two different-flavor SM leptons in a four-fermion vertex, we obtain the effective lepton-flavor violating Yukawa shown in the left panel of Fig.~\ref{Fig:basicDiagrams}. This arrangement avoids tree-level LFV couplings between the Higgs and SM leptons, but comes at the price of introducing a cutoff scale $\Lambda$. Technically, our setup is an instance of the 2HDM where in the UV theory the four-fermion vertex resolves into a renormalizable interaction mediated by an extra scalar $\phi$. This extra doublet Higgs field has squared mass $\Lambda^{2}$ and gets a vev through mixing with the light Higgs in order to misalign the Yukawa and mass bases. The vector-like fermions with sizable couplings to both Higgses are the key ingredient, without them any LFV will  be proportional to the lepton Yukawa coupling and the lepton mass making this effect completely negligible. It is through the sizable couplings of this amplitude, the extra fermion masses and the cutoff scale $\Lambda$ that we intend to generate a $h\rightarrow \tau \mu$ branching fraction at the value (\ref{BFcombined}).

\begin{figure}[h!]
\centering
\includegraphics[scale=0.5]{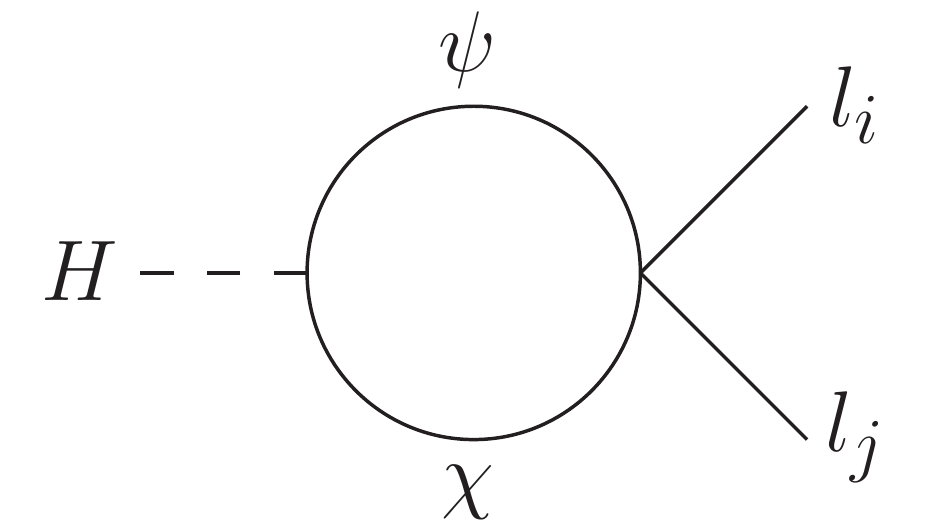}
\hspace{10mm}
\includegraphics[scale=0.5]{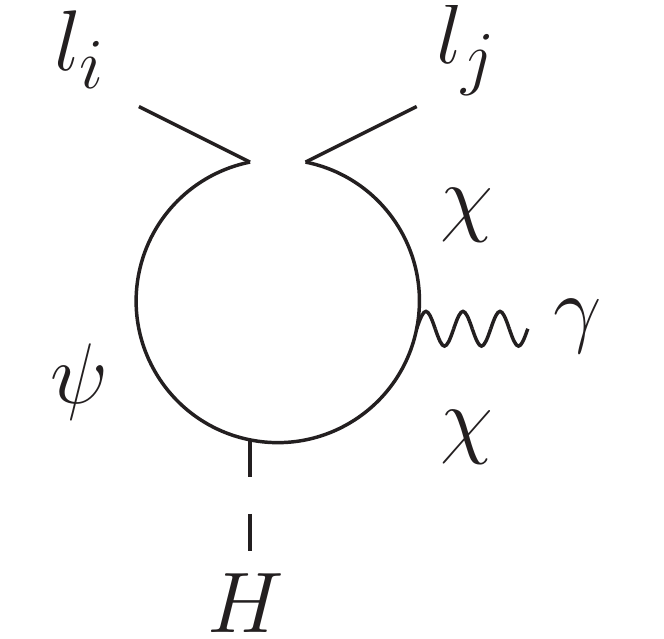}
\caption{\textbf{Left:} Effective flavor off-diagonal Yukawa coupling for $h\rightarrow l_{i}l_{j}$ through exchange of vectorlike fermions $\psi,\chi$. \textbf{Right:} induced amplitude for $l_{i} \rightarrow l_{j} \gamma$ via an open fermion line.}
\label{Fig:basicDiagrams}
\end{figure}

In Eq.~(\ref{LagrangianYukawa}), gauge invariance dictates that one of the new fermions must be a $SU(2)_{\text{L}}$ doublet and the other one a singlet, and the hypercharge must satisfy $Y_{\psi} - Y_{\chi} = 1/2$. This leaves some freedom in the overall hypercharge of the vectorlike matter\footnote{As $\psi, \chi$ are vectorlike, there is no constraint from anomaly cancellation.}, i.e $\psi=(\boldsymbol{2},1/2+x),\,\chi=(\boldsymbol{1},x)$ for any value of $x$. By exploiting this freedom in the overall hypercharge, we can control the interactions between $\psi, \chi$ and SM matter. Specifically, to obtain the simplest setup with the properties we are interested, we make the following considerations: (i) we take $x \ne 0$ to avoid introducing a Majorana fermion $\chi$ (and the Majorana masses that is then allowed); (ii) in general, fermions with non-integer electric charges are stable, so we restrict $x$ to be an integer; (iii) if $x=-1$, the new fermions carry the same quantum number as SM leptons and one could write Yukawa interactions (or dimension-3 mass mixings) between $\psi, \chi$ and the SM. These renormalizable interactions are undesirable for a couple of reasons. First, additional interactions means a more complicated setup. More problematic, chiral-vector fermion mixing (also allowed if $x=+1$) will induce both LFV Higgs decays and LFV radiative decays, and the restrictive parameter space of the latter makes the former incompatible with the rate in Eq.~(\ref{BFcombined}) \cite{Falkowski:2013jya}; (iv) finally, some values of $x$ allow less operators than others. For instance, another effect of $x=-1$ is that it forces us to write down five operators, which requires  introducing more free parameters (but no different effects). In order to avoid these complications, we choose $x=2$, under which only the three operators in Eq.~(\ref{Lagrangian4f}) appear (details in Appendix \ref{sec:AppendixChoice}).

Having fully specified the quantum numbers of $\psi,\, \chi$, we now must ask if there are other operators of equal or lesser mass dimension we need to include, i.e. other operators that can contribute to LFV Higgs or radiative decays. For $x=2$, there is one other dimension-6 term, $H^{2}H^{\dag}\psi^{c} \chi$ (and its charge-conjugate) which we can use to build a $h\rightarrow l_{i}l_{j}$ diagram like the one in Fig.~\ref{Fig:basicDiagrams} but with two more Higgs vev insertions at the Yukawa vertex. However, contributions from this operator are suppressed by $v^{2}/\Lambda^{2}$ compared to the term already included in Eq.~(\ref{LagrangianYukawa}). More importantly, although $x=2$ does not allow the new fermions to decay via renormalizable couplings to the SM, higher-order operators such as $e^{c\dag}e^{c\dag}L\psi$ do permit the decay of the lightest component of the fermion doublet $\psi$ (mass $M$) with lifetime $\tau_{\psi}^{-1} \sim M^{5}/\Lambda^{4}$. For the parameter space we will be interested in, the lifetime of the lighter state is sufficiently short to avoid any issues.

Returning to the four-fermion vertices appearing in Eq.~(\ref{Lagrangian4f}), there are two different ways to contract the Lorentz indices. The differences can be seen clearly when we try to close the loop of heavy fermions in Fig.~\ref{Fig:basicDiagrams}: either the (Lorentz) indices of the SM leptons and the vectorlike fermions contract among themselves separately, leaving a closed fermion loop, or each SM lepton contracts with one of the internal fermions, in which case there is an open fermion line. Both types of contractions contribute to $h\rightarrow l_{i}l_{j}$. The $l_{i}\rightarrow l_{j}\gamma$ amplitude arises by attaching photon lines to $\psi$ or $\chi$ (right diagram of Fig.~\ref{Fig:basicDiagrams}), and, due to the trace over the $\gamma$-matrix structure in the internal loop, only the open-line diagram is able to generate the coefficient $\sigma^{\mu \nu}$ of the dipole operator needed for $l_{i} \rightarrow l_{j} \gamma$. The width expressions for these two processes are:
\begin{align}
\Gamma(h\rightarrow l_{i}l_{j})
&= \dfrac{m_{h}}{16\pi}\left( \dfrac{M_{\chi}}{16\pi^{2}\Lambda^2} \right)^{2}\biggl[ \biggl( y^{2}(\lambda_{2})_{ij}^{2}+y^{'2}(\lambda_{3})_{ij}^{2} \biggr)\mathcal{H}_{\text{closed}}^{2}+(\lambda_{1})_{ij}^{2}y^{2}\mathcal{H}_{\text{open}}^{2} \biggr], \label{WidthhTauMu} \\
& \notag \\
\Gamma(l_{i} \rightarrow l_{j} \gamma)
&= \dfrac{m_{l_{i}}^{3}}{16\pi}\left( \dfrac{1}{16\pi^{2}} \right)^{2}\left[ \dfrac{veQ_{\chi,\psi}}{\Lambda^{2}} \right]^{2} 2(\lambda_{1})_{ij}^{2}y^{2}\biggl[ \mathcal{G}^{2}(M_{\psi},M_{\chi})+\mathcal{G}^{2}(M_{\chi},M_{\psi}) \biggr]. \label{WidthTauMuGamma}
\end{align}
The loop functions $\mathcal{H}_{\text{open,closed}}$ go approximately as $(M_{\psi}^{2}+M_{\chi}^{2})/M_{\chi}^{2}$, and $\mathcal{G}$ as $(M_{\chi}^{6}+M_{\chi}^{6})/(M_{\chi}^{2}-M_{\psi}^{2})^{3}$ (for the exact forms, see Appendix~\ref{sec:AppendixFunctions}). In the regions of interest, $|\mathcal{H}_{\text{closed}}|$ and $|\mathcal{H}_{\text{open}}|$ vary between 1 and 2, while $|\mathcal{G}|\approx 1$. 

In Eq.~(\ref{WidthTauMuGamma}), we see that $(\lambda_{1})_{ij}$ is the sole coupling governing $l_{i}\rightarrow l_{j}\gamma$, while all three $(\lambda_{k})_{ij}$ contribute to $h \rightarrow l_{i}l_{j}$. In principle, this means one could switch off $(\lambda_1)_{\mu\tau}$ while maintaining $(\lambda_{2})_{\mu\tau}, (\lambda_3)_{\mu\tau} \ne 0$, thereby reproducing $h\rightarrow \tau \mu$ without radiative LFV decays. However, this choice is somewhat tuned, as it implies a particular choice of UV boundary conditions for our effective field theory. Moreover, as we will show in Sec.~\ref{sec:EffectsLambda1Eff}, $(\lambda_1)_{ij}$ can be induced via loops involving $(\lambda_2)_{ij}$ and $(\lambda_3)_{ij}$, so the choice $(\lambda_{1})_{ij}=0$ is not radiatively stable.

The LFV rates in our model are described by Eqs.(\ref{WidthhTauMu}) and (\ref{WidthTauMuGamma}), and apply equally to each pair of charged leptons $l_{i}$ and $l_{j}$ with $i\neq j$. For simplicity, we set $(\lambda_{k})_{ij} = (\lambda_{k})_{ji}$. The parameter space relevant to LFV between any two sectors is given by $\bigl( y, y', (\lambda_{k})_{ij}, M_{\psi}, M_{\chi}, \Lambda \bigr)$, though we will adopt the equivalent set $\bigl(y$, $y'/y$, $(\lambda_{k})_{ij}$, $M_{\text{fermion}}$, $M_{\chi}/M_{\psi}$, $\Lambda \bigr)$ -- where $M_{\psi}$ is renamed $M_{\text{fermion}}$ -- for convenience. Although not a small list of parameters, each observable is sensitive only to a subset of this list. As a first step, we take the following inputs:
\begin{equation}
y'/y=1,~~~~~(\lambda_{1})_{ij}=(\lambda_{2})_{ij}=(\lambda_{3})_{ij},~~~~~M_{\chi}/M_{\psi}=1. \label{input1}
\end{equation}

The strategy we follow when extracting numerical results in the next section is as follows: we first uncover the $(\Lambda,M_{\text{fermion}})$ parameter space where the LFV Higgs and radiative decays are compatible with each other. Next, we move to the $\mu-e$ sector, though in order to place the $\tau-\mu$ and $\mu-e$ constraints on the same plane we will need to make additional assumptions on the relations between the $(\lambda_i)_{\tau\mu}$ and $(\lambda_i)_{\mu e}$.

Before moving to our numerical results, we emphasize that the details of the UV completion and underlying flavor texture origin are beyond the scope of this paper, hence there will be no top-down bias when selecting points in the space of free parameters. In particular, we do not introduce any assumption that could force the coupling $(\lambda_{1})_{ij}$, which parametrizes $l_{i}\rightarrow l_{j}\gamma$, to vanish. Such assumptions could be the result of discrete symmetries between different lepton families, or continuous symmetry like $U(1)_{L_i-L_j}$. Brief comments on the numerical 
effects of $(\lambda_{1})_{ij}=0$ are presented in Sec.~\ref{sec:EffectsLambda1Eff} only after having analyzed the generic cases in which no $(\lambda_{1})_{ij}$ vanishes.

\section{Numerical results}   
\label{sec:Numerical}

\subsection{Constraints by $\tau \rightarrow \mu \gamma$}
\label{subsec:TauMuGamma}

For simplicity, in this section we will drop the flavor label in the $\mu-\tau$ couplings and write them as $\lambda_{i}$. In our effective theory approach, it is preferred to keep the $\lambda_{i}$ couplings with size $\mathcal{O}(1)$, otherwise the meaning of $\Lambda$ becomes murky. As mentioned earlier, without specific knowledge of the UV physics that completes Eq.~(\ref{Lagrangian4f}), setting just one of the $\lambda_{i}$ to zero looks tuned, so we will work with $\lambda_{i}=1$ for $i=1,2,3$. Taking the benchmark point $y = 0.25$, the region in $(\Lambda, M_{\text{fermion}})$ parameter space that accommodates the observed $h\rightarrow \tau \mu$ branching fraction is shown below in Fig.~\ref{Fig:TauMuSector}, along with the regions excluded by BaBar and Belle limits on $\tau \rightarrow \mu \gamma$~\cite{Aubert:2009ag,Hayasaka:2007vc} ($\mathcal{B}r(\tau \rightarrow \mu \gamma)< 4.4\times 10^{-8}$). Setting $y'=y$ implies that the $\mathcal{B}r(h\rightarrow l_{i}l_{j})$ is proportional to $y$, so the $h\rightarrow \tau \mu$ rate appears as a straight line  in the $(\Lambda,M_{\text{fermion}})$ plane; as the value of $y = y'$ is lowered (raised), the slope of the best fit line increases (decreases). Evidently, both observables are compatible in much of the selected $(\Lambda,M_{\text{fermion}})$ window. We have verified that slight deviations from $M_{\chi}/M_{\psi}=1$ do not alter the best fit band or the contour significantly\footnote{Larger deviations from $M_{\chi}/M_{\psi}=1$ are not considered because consistency requires having a fermion mass splitting much smaller than the difference between these and $\Lambda$.}. In the neighborhood of the dip feature (to the left of the bump) the ratio $M_{\text{fermion}}/\Lambda$ approaches the value that minimizes the loop function $\mathcal{G}$ that sets the $\tau \rightarrow \mu \gamma$ contour.

\begin{figure}[h!]
\centering
\includegraphics[scale=0.59,keepaspectratio=true]{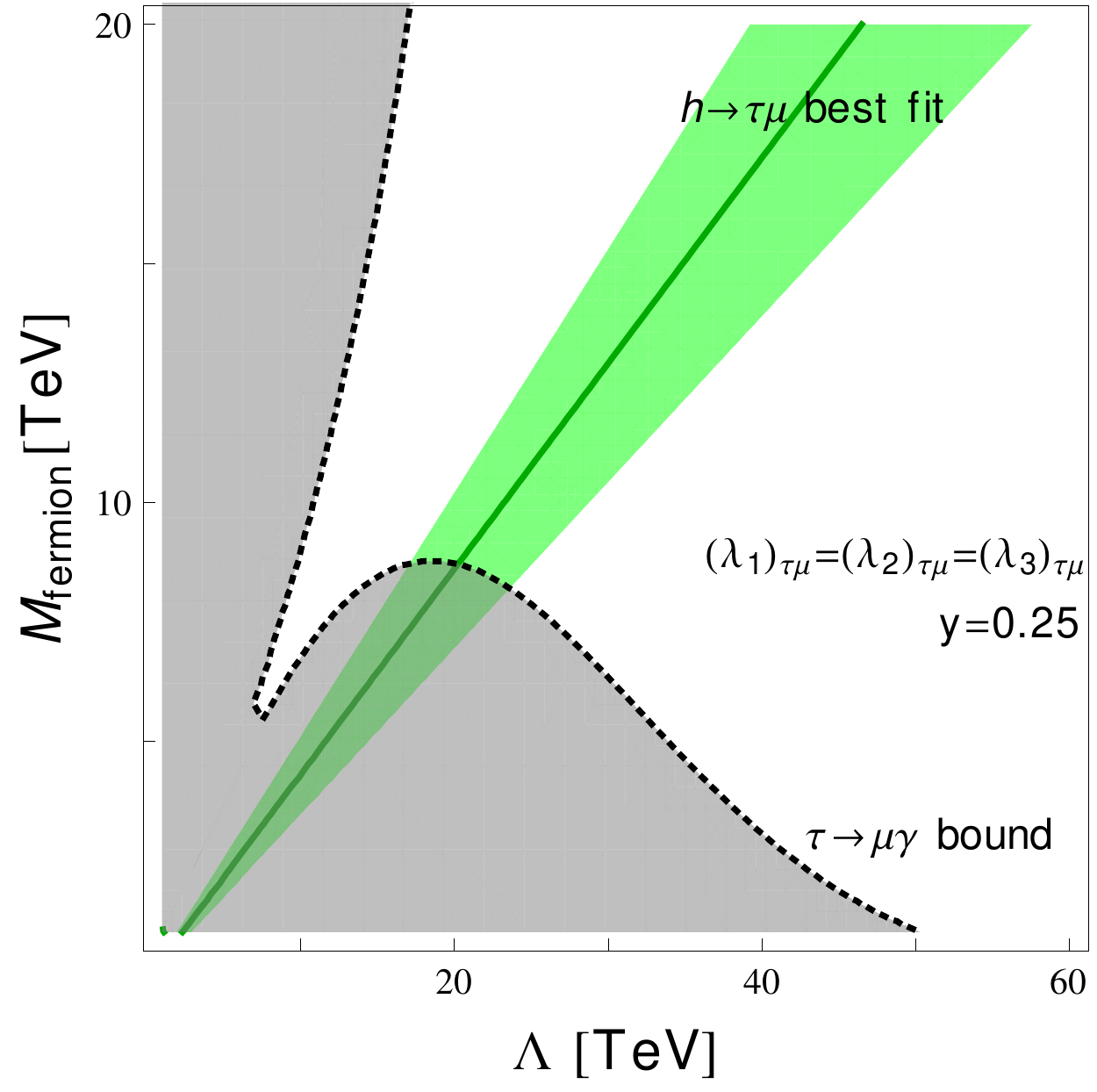}
\caption{The $2.6\,\sigma$ LFV Higgs decay best fit (green band) and the bound on the lepton radiative decay (dotted line) in the $\tau-\mu$ sector, with $y'=y=0.25$ and the $\mu\tau$ couplings set to $(\lambda_1)_{\tau\mu}=(\lambda_{2})_{\tau\mu}=(\lambda_{3})_{\tau\mu}=1$. Near the dip feature, $M_{\text{fermion}}/\Lambda$ approaches the value that minimizes the loop function $\mathcal{G}$.}
\label{Fig:TauMuSector}
\end{figure}

Even though our focus is on the off-diagonal (hence LFV) flavor entries of the $\lambda_{i}$ matrices, quantum corrections would generate diagonal $(\lambda_{k})_{ii}$, even if these couplings happened to vanish at a certain scale. So let's instead imagine that we have them around right from the start.
Flavor-diagonal $(\lambda_{k})_{ii}$ couplings will correct the SM fermion Yukawas through a four-fermion loop. A diagonal coupling $\lambda_{\tau \tau}$ of the same size as $\lambda_{\tau \mu}$ (equal to 1 in our benchmark) corrects the tau Yukawa by an amount $(1/16\pi^{2})(y\lambda_{\tau \tau}/\Lambda^{2})M_{\text{fermion}}^{2}$. This correction, when evaluated at phenomenologically acceptable points $(\Lambda,M_{\text{fermion}})$ near the cusp in Fig.~\ref{Fig:TauMuSector}, is one order of magnitude smaller than the tau Yukawa, therefore it is a sub-leading effect. In the case of the muon and the electron, order-one couplings $\lambda_{\mu \mu}$ and $\lambda_{ee}$ cannot be used as their corrections to the corresponding Yukawas are too large at the same $(\Lambda,M_{\text{fermion}})$ values.

If the flavor pattern of the UV theory behind our setup somehow yields a texture where the only off-diagonal entries in the $\lambda_i$ are the $\mu-\tau$ entries, then there are no other constraints to consider\footnote{There are constraints from LFV decays $\tau \rightarrow 3\mu$, however these are far weaker than the bounds from $\tau \rightarrow \mu \gamma$, as we will review shortly.}. However, such a UV flavor structure seems rather ad hoc, so we would like to expand our setup to broader flavor textures. Specifically, we will now allow other off-diagonal entries in the $\lambda_i$ matrices and ask what values $(\lambda_{i})_{\mu e}$, $(\lambda_{i})_{\tau e}$, etc. are allowed (relative to $(\lambda_{i})_{\tau\mu}= 1$, and assuming the new physics scale $\Lambda$ is fixed). Within this framework, the tightest LFV constraints come from the radiative decay $\mu \rightarrow e \gamma$, so we turn to this process next.

\subsection{Constraints by $\mu \rightarrow e\gamma$}
\label{subsec:MuEGamma}

The form of $\Gamma(\mu \rightarrow e\, \gamma)$ is given by Eq.~(\ref{WidthTauMuGamma}), with $m_{l_i} = m_{\mu}$ and $(\lambda_1)_{ij} = (\lambda_1)_{\mu e}$, and the strongest limit on $\mu \rightarrow e\gamma$ currently comes from the MEG experiment~\cite{Adam:2013mnn}, $\mathcal{B}r(\mu \rightarrow e\gamma)<5.7\times 10^{-13}$. In order to discern the $(\Lambda,M_{\text{fermion}})$ regions permitted by $\mu \rightarrow e\, \gamma$ on top of the $\tau-\mu$ observables, we vary $\lambda_{\mu e}$ for fixed $\lambda_{i} = 1$ (the $\tau-\mu$ couplings) until its allowed region overlaps with the best fit band; see Fig.~\ref{Fig:lambdaMuE} below.  

\begin{figure}[h!]
\centering
\includegraphics[scale=0.59,keepaspectratio=true]{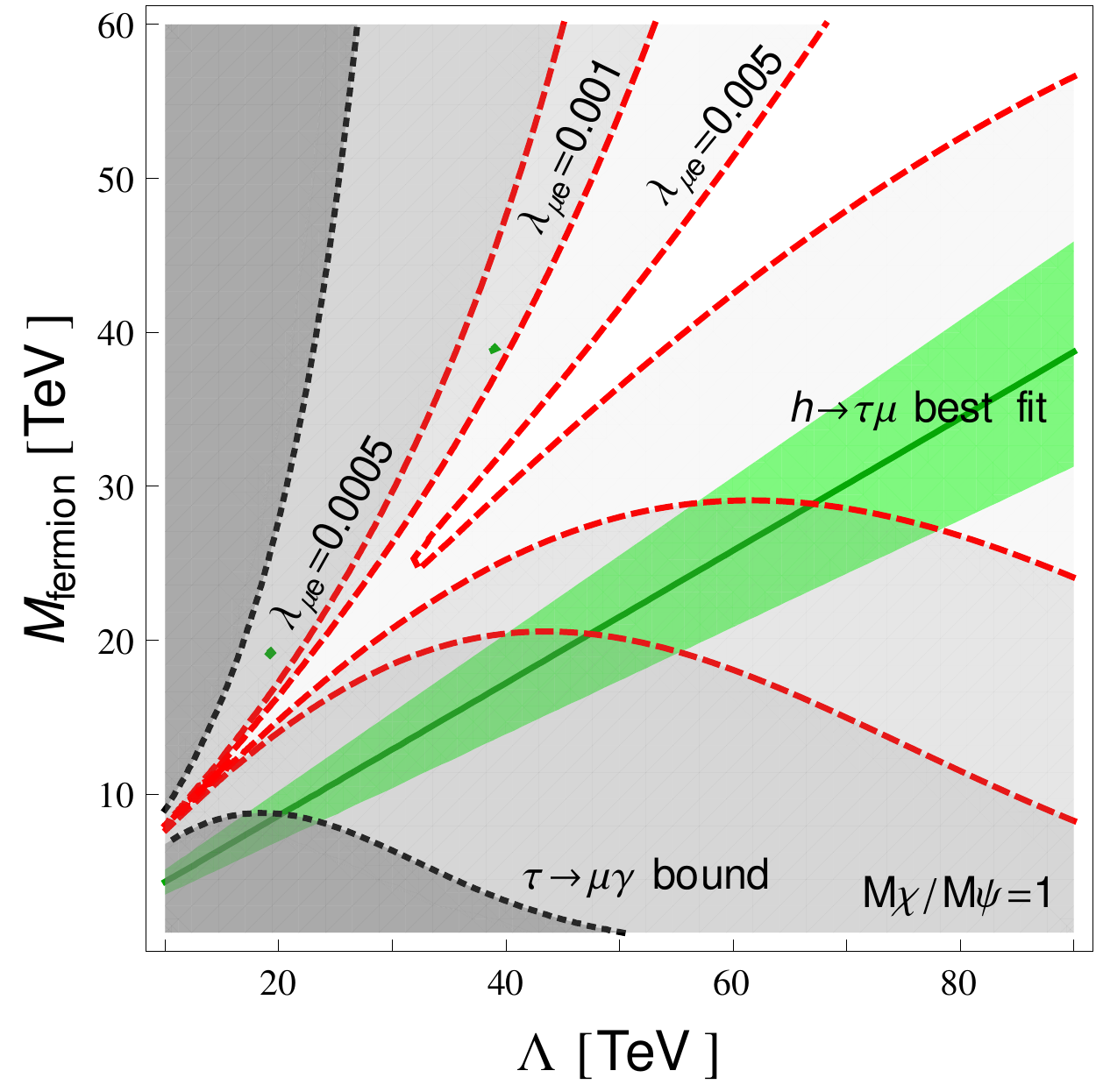}
\includegraphics[scale=0.59,keepaspectratio=true]{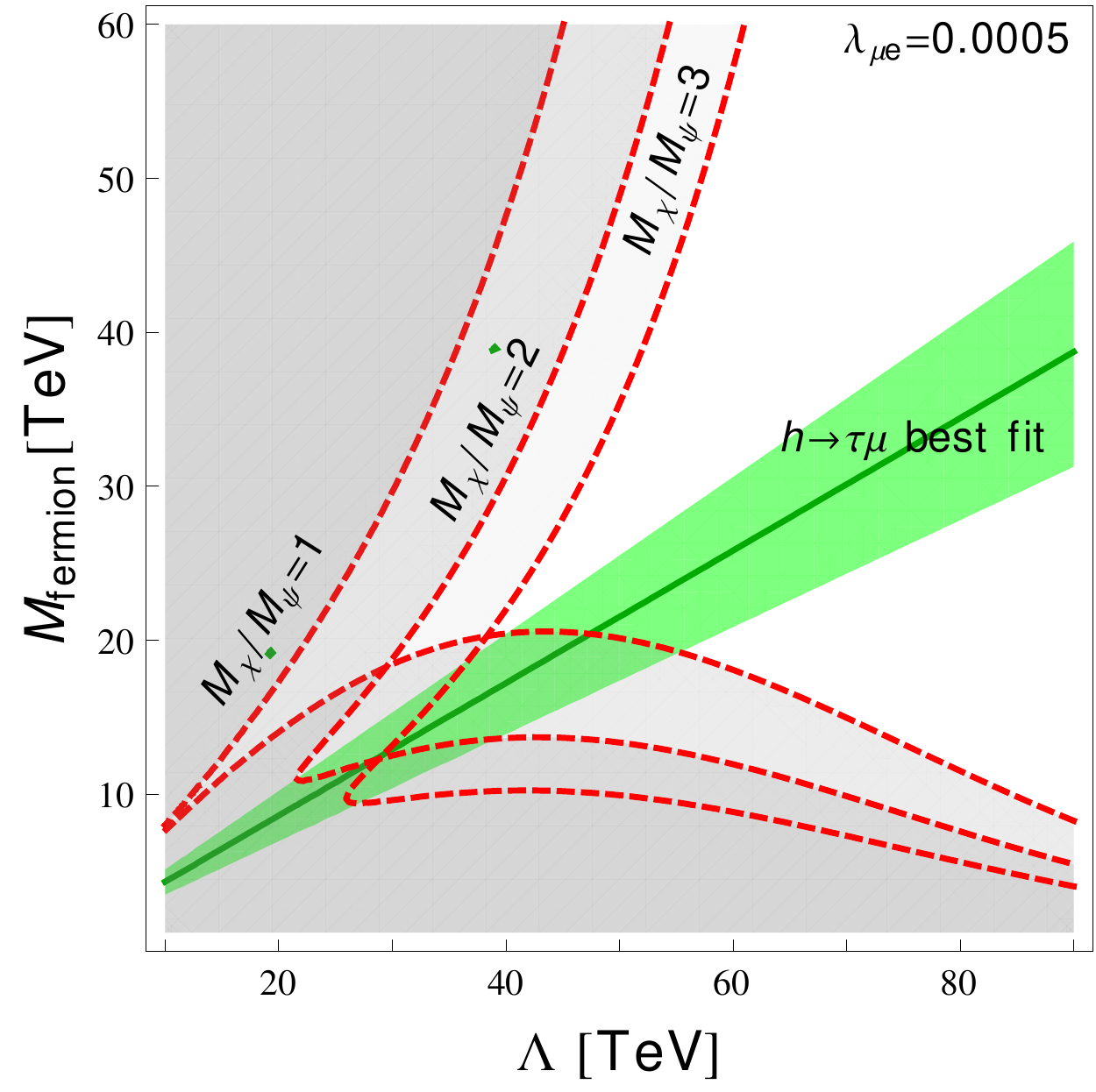}
\caption{\textbf{Left:} Radiative decay $\tau \rightarrow \mu \gamma$ (dotted line) and contours of $\mu \rightarrow e\gamma$ with varying $\lambda_{\mu e}$ (dashed red lines) and fixed fermion mass splitting. For each $\lambda_{\mu e}$, the forbidden regions (shadowed) are to the left and below the countours. \textbf{Right:} Contours of $\mu \rightarrow e\gamma$ for various fermion mass splittings (red dashed lines)  with fixed $\lambda_{\mu e}=0.0005$. In both panels $y'=y=0.25$ and $\lambda_{1}=\lambda_{2}=\lambda_{3}=1$.}
\label{Fig:lambdaMuE}
\end{figure}

In the left panel, contours of the $\mathcal{B}r(\mu \rightarrow e\gamma)$ (red dashed lines) and $\mathcal{B}r(\tau \rightarrow \mu \gamma)$ (black dotted line) indicate bounds for the same input choices as $h\rightarrow \tau \mu$ and several values of $\lambda_{\mu e}$ values, superimposed over Fig.~\ref{Fig:TauMuSector} (using a larger window than in that graph). The area below the red (black) contour has $\mu \rightarrow e\gamma$ ($\tau \rightarrow \mu \gamma$) rates larger than the bound, thus those points are forbidden and shaded in light (dark) gray. The overlap of the best fit band with the white safe region happens for $\lambda_{\mu e}=0.0005$ near $\Lambda \approx 50\text{ TeV}$ and $M_{\text{fermion}}$ as low as $20\text{ TeV}$.

In the right panel of Fig.~\ref{Fig:lambdaMuE}, we show how the viable parameter $(\Lambda, M_{\text{fermion}})$ space shifts as we vary the ratio of the two vectorlike fermion masses $M_{\chi}/M_{\psi}$ (here, $M_{\text{fermion}}$ stands for the lighter of $M_{\psi,\chi}$) while keeping $\lambda_{\mu e} = 0.0005$ fixed. As the ratio of $M_{\chi}/M_{\psi}$ increases, the allowed parameter region slides to higher $\Lambda, M_{\text{fermion}}$\footnote{Again, we stick to $M_{\chi}/M_{\psi}$ near to 1 to keep the splitting small compared to $M_{\text{fermion}}/\Lambda$.}. Recalling that the slope of the $h\rightarrow \tau \mu$ band is controlled by the Yukawas,\footnote{The $\mu \rightarrow e\gamma$ contour also changes by varying $y$, $y'$ but its effect is less pronounced than when varying $\lambda_{\mu e}$.} there is freedom to place the crossing of the allowed regions over a wide range of $\Lambda$ values (once $M_{\text{fermion}}$ has been fixed). However, for a fixed $\mu \rightarrow e\gamma$ contour, smaller values of $y,y'$ (for which the slope of the $h\rightarrow \tau \mu$ band is larger and the allowed $\Lambda$ is smaller) are not preferred as they bring $M_{\text{fermion}}$ increasingly closer to $\Lambda$ and therefore into a regime where our effective field theory is less reliable.

In Fig.~\ref{Fig:zoomBeforeAfter} below, we zoom in on two example regimes where $\mathcal{B}r(h \rightarrow \tau\mu)$ matches observation and all LFV $\ell_i \rightarrow \ell_j\, \gamma$ constraints are satisfied. In both examples, $(\lambda_1)_{\mu e} = 0.0005$ and $(\lambda_{i})_{\tau\mu} = 1$:
\begin{itemize}
\item{$\Lambda\approx 35$ TeV \textit{regime}:} Here $y$ is decreased to $0.17$ and $y' \ne y$ (both Yukawas are still perturbative). For equal $(\lambda_i)_{\tau \mu}$ couplings and equal fermion masses, the allowed region is around $\Lambda=35\text{  TeV}$ and $M_{\text{fermion}}=15 \text{ TeV}$ (left panel in Fig.~\ref{Fig:zoomBeforeAfter}).
\item{$\Lambda\approx 80$ TeV \textit{regime:}} Here $y, y'$ are larger but equal, $y=y'=0.4$, all $(\lambda_{i})_{\mu e}$ are equal, as are the fermion masses. For this choice, the allowed region moves all the way up to the neighborhood of a 80-TeV $\Lambda$ (right panel in Fig.~\ref{Fig:zoomBeforeAfter}) with $M_\text{fermion}=20\text{ TeV}$.
\end{itemize}

We remind the reader that the $\mu \rightarrow e\gamma$ bound only constrains $(\lambda_1)_{\mu  e}$ and not $(\lambda_2)_{\mu  e}$ or $(\lambda_3)_{\mu  e}$. The latter two couplings certainly participate in other LFV amplitudes in the $\mu-e$ sector, for example in Higgs-mediated $\mu \rightarrow eee$ and nuclear $(\mu-e)$ conversion, but when these processes are studied in Sec.~\ref{subsec:MuEGamma} we will see that no meaningful bound on the $(\lambda_{2,3})_{\mu e}$ can be extracted from them since the $(\lambda_{2,3})_{\mu e}$ dependence is accompanied by factors of $\alpha_{em}$ and/or lepton Yukawa couplings. However, if we follow the same effective field theory logic we invoked in the $\tau-\mu$ LFV -- that the three $\lambda_i$ should be similar in size -- we can already extend the $(\lambda_1)_{\mu e}$ bound to the entire $\mu-e$ sector of our model.

 In the analysis above, we found that a ratio $\lambda_{\mu e}/\lambda_{\tau \mu}\sim 10^{-4}$ is needed in order to meet the experimental bounds. This value is comparable to the ratio $y_{e}/y_{\tau}$, and it may be taken as a hint of the underlying flavor physics at the cutoff $\Lambda$. Said another way, the size of $\lambda_{\mu e}/\lambda_{\tau \mu}$ indicates that the tradeoff to satisfy LFV bounds, at least in the $\mu-e$ sector, is to abandon the possibility of having all four-fermion couplings of order 1. However, if we impose hierarchy among couplings by demanding $(\lambda_{k})_{ij} \propto y_i y_j$ (here $y_{i}$ are the lepton Yukawa couplings), the scale $\Lambda$ is so  dramatically reduced that $h \to \tau \mu$ and $\tau \to \mu \gamma$ are no longer compatible. Perhaps a better approach is to model-build the hierarchy in the $\lambda_{i}$ by generating the four-fermion operators in Eq.~(\ref{Lagrangian4f}) for different SM generations at different scales.

Before finishing this section, we comment on the size of the correction to $\lambda_{\mu e}$ generated by the four-fermion vertex $\lambda_{\tau \mu}$ when the tau turns into an electron leg through a neutrino-W-boson loop. This LFV self-energy will be proportional to sum of the squared neutrino masses (times the Fermi constant times neutrino mixing entries) which can be overestimated by the bound on the sum of the three absolute neutrino masses~\cite{Agashe:2014kda}. Since these are $\mathcal{O}(\text{eV})$ by themselves, a negligible factor $10^{-23}$ is expected when compared to the leading order $\lambda_{\mu e}=0.0005$. Therefore, the four-fermion couplings are radiatively stable.

\begin{figure}[h!]
\centering
\includegraphics[scale=0.59,keepaspectratio=true]{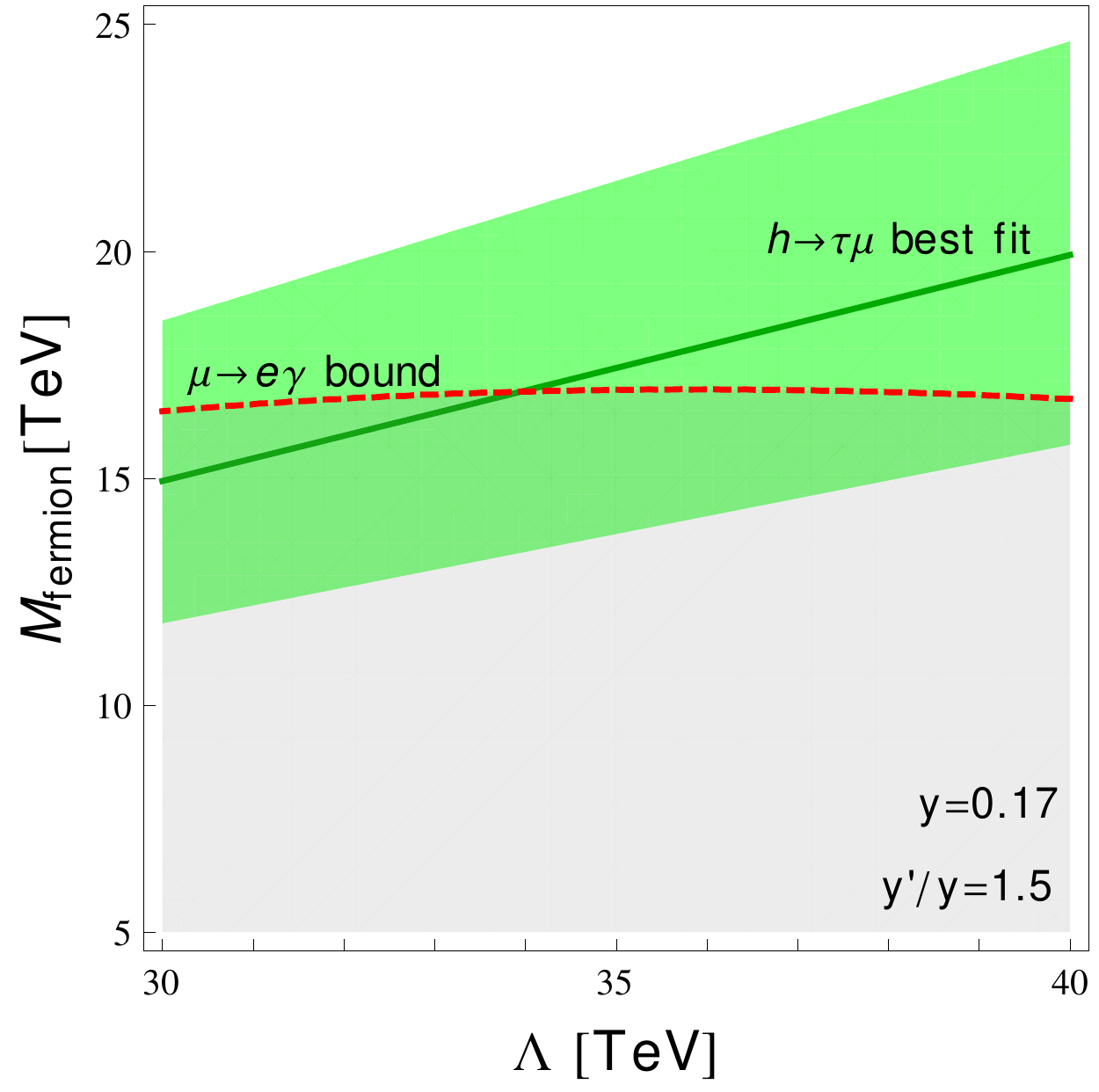}
\includegraphics[scale=0.59,keepaspectratio=true]{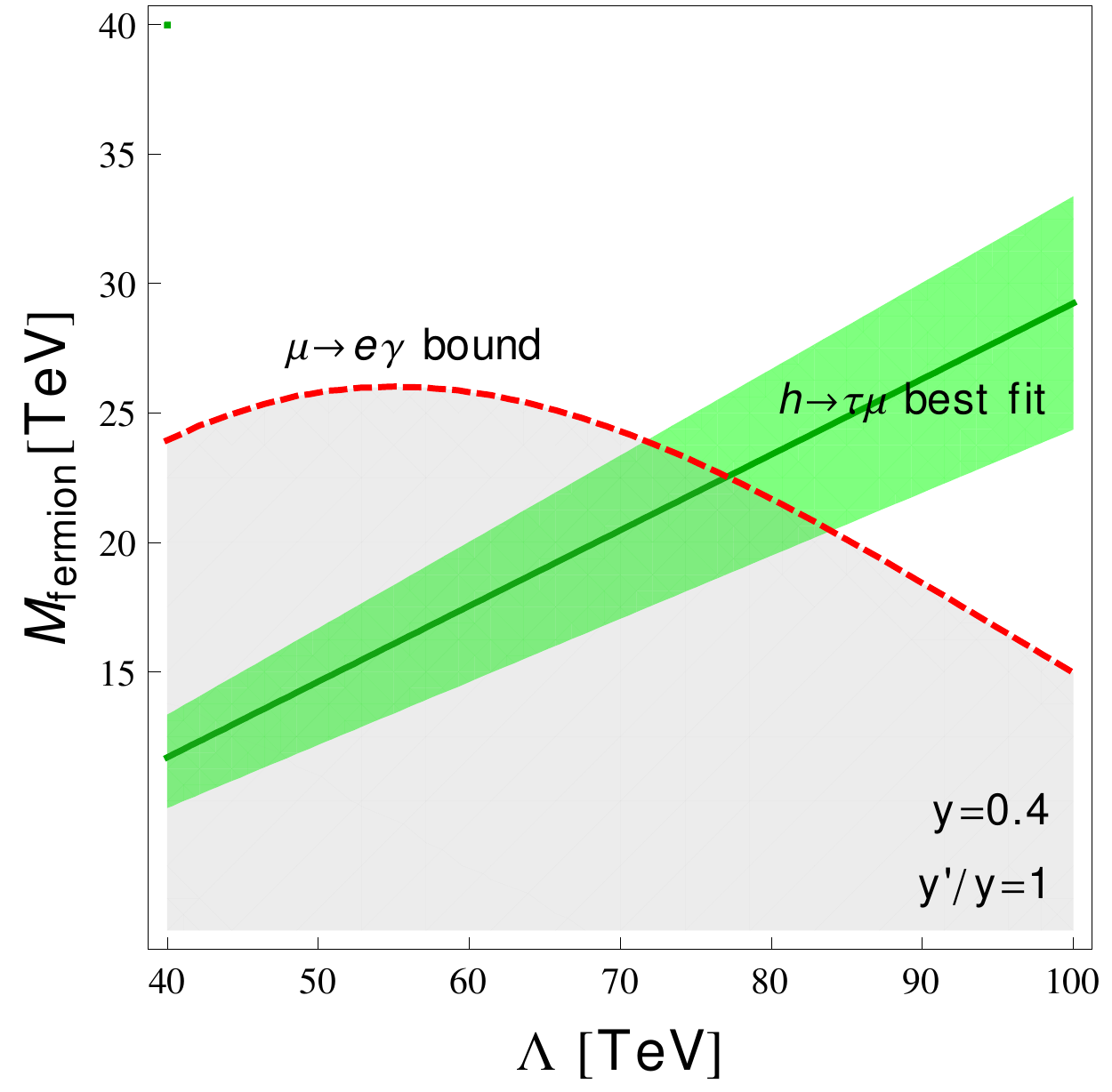}
\caption{\textbf{Left(right) panel:} scenario for a low (high) $\Lambda$ scale obtained by lowering(increasing) the Yukawa couplings. In both panels $\lambda_{\mu e}=0.0005$, $M_{\chi}/M_{\psi}=1$ and the $\tau-\mu$ couplings fixed at $\lambda_{1}=\lambda_{2}=\lambda_{3}=1$.}
\label{Fig:zoomBeforeAfter}
\end{figure}


\subsection{Effects of $(\lambda_{1})_{ij}=0$}
\label{sec:EffectsLambda1Eff}

In previous sections, we pointed out that when $\lambda_{1}$ is set to zero, the Higgs decays $h\to l_i l_j$ are not constrained by the radiative processes $l_i \to l_j \gamma$. However, even in the hypothetical scenario where $\lambda_{1}=0$ at tree-level, one loop-corrections to the four-fermion vertices $\lambda_{2}$ and $\lambda_{3}$ induce an effective $\lambda_{1}$ coupling. Specifically, $\gamma$ and $Z$ exchange between two uncontracted fermions in the $\lambda_2$, $\lambda_3$ terms in Eq.~(\ref{Lagrangian4f}) generates the Lorentz structures required for $\lambda_1$. To a good approximation, the effective $\lambda_1$ coming from these loop effects is $\lambda_{1}^{\text{eff}} \sim (\alpha_{em}/4\pi)\cdot2\,(\lambda_{2}+\lambda_{3})$. We now discuss how our bounds change if $\lambda_1$ -- either in the $\tau-\mu$ sector or in the $\mu - e$ sector -- is reduced to this loop-level value:

\begin{itemize}

\item{\textit{Effects on $h \rightarrow \tau \mu$}:}

Numerically, the change in the slope of the best fit band is sub-percent level, so there is no significant consequences to the analysis of $h\rightarrow \tau\mu$. This is explained by the fact that the $\lambda_{1}$ part of $\Gamma(h\rightarrow \tau \mu)$ was already subdominant with respect to the terms with $(\lambda_{2,3})_{\tau \mu}$ due to $\mathcal{H}_{\text{open}}^{2}\approx (1/4)\mathcal{H}_{\text{closed}}^{2}$. 

\item{\textit{Effects on $\tau \rightarrow \mu \gamma$}:}

Reducing $\lambda_1$ to the loop-induced value, the region forbidden by $\tau \to \mu \gamma$ loosens enough to practically disappear from the analyzed range in the $(\Lambda,M_{\text{fermion}})$ plane. The relaxed bounds in this circumstance means there is a sliver of parameter space, where the $\psi ,\chi$ fermions are accessible at the LHC. As one concrete example, for $y/(4\pi) \sim 0.3, \lambda_2 = \lambda_3 = 1$, we find that $M_{\text{fermion}} \sim 600$ GeV and $\Lambda \sim 10$ TeV.

\item{\textit{Effects on $\mu \rightarrow  e\gamma$ and constraints from $h \rightarrow \mu e$}:}

In our earlier analysis, we found  $(\lambda_{1})_{\mu e}$ needed to be $\mathcal{O}(\sim 10^{-4}\, \lambda_{\tau\mu})$ for there to be an overlap between the $\mu \to e\, \gamma$, $h \to \tau \mu$ and $\tau \to \mu \gamma$ allowed regions, as we extended this bound to $(\lambda_{2})_{\mu e},(\lambda_{3})_{\mu e}$ purely based on the vague effective field theory argument that couplings involving the same fields with the same mass dimension should be the same order of magnitude. Now, we can ask a different question: if we assume that $(\lambda_1)_{\mu e}$ is solely generated by loops, how do the bounds on $\mu \to e\,\gamma$ translate into bounds on $(\lambda_{2,3})_{\mu e}$, and how do those bounds compare to bounds coming from the Higgs decay $h \to \mu e$, a process that is sensitive (see Eq.~(\ref{WidthTauMuGamma})) to all three $(\lambda_i)_{\mu e}$? To answer the first part of the question, the $\mu \to e\, \gamma$ bounds are shown below assuming $(\lambda_1)_{\mu e}$ = $(\lambda^{\text{eff}}_1)_{\mu e}$ for three different values of $(\lambda_{2,3})_{\mu e}$. The $(\lambda_2)_{\mu e} = (\lambda_3)_{\mu e} =  0.2$ contour roughly corresponds to $(\lambda^{\text{eff}}_1)_{\mu e} \sim 0.0005$.

\begin{figure}[h!]
\centering
\includegraphics[scale=0.63,keepaspectratio=true]{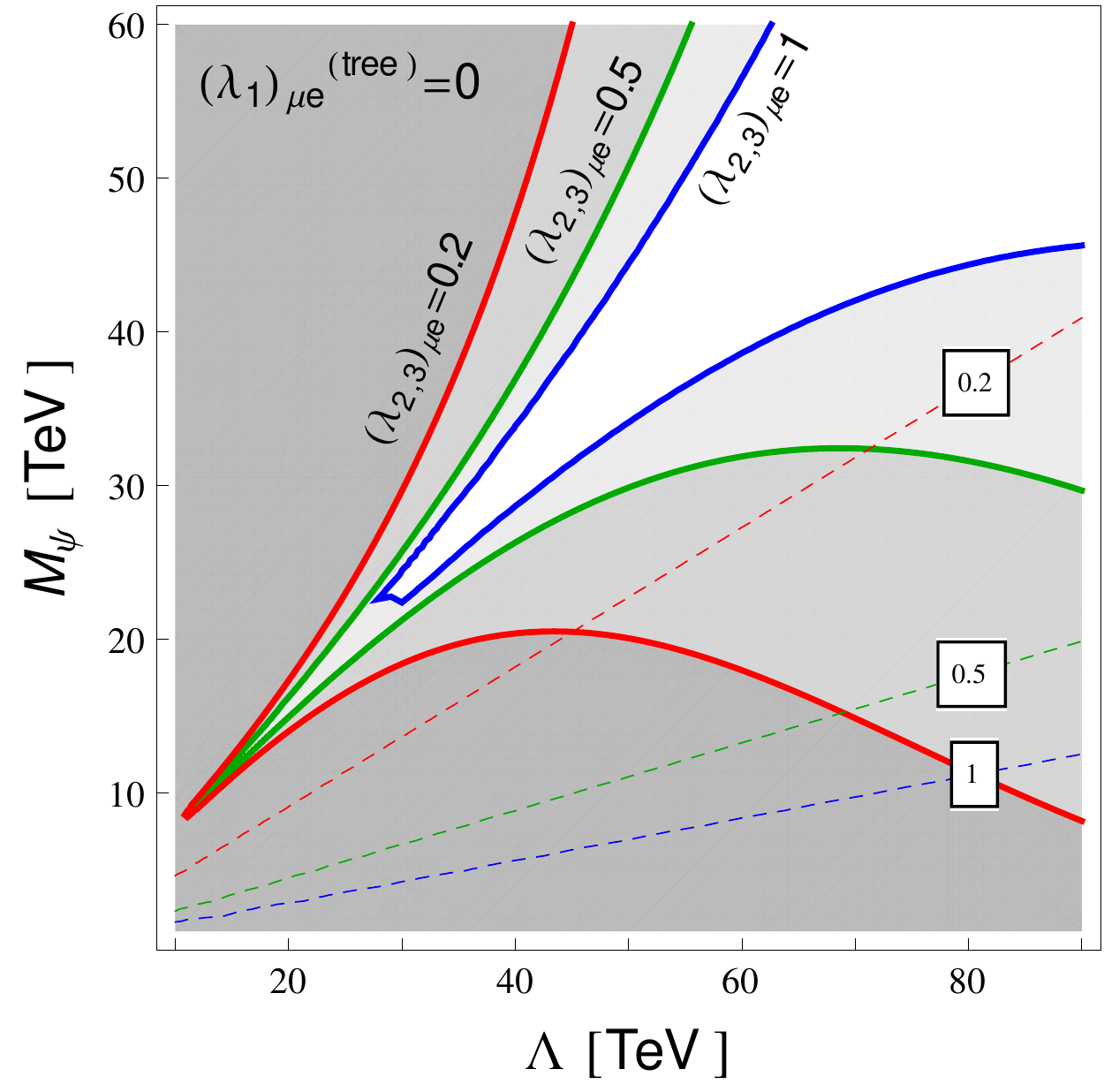}
\caption{Solid lines show the contours from the bound on $\mu \rightarrow e\gamma$ taking $(\lambda_{1})_{\mu e} =  (\lambda^{\text{eff}}_{1})_{\mu e}$, the value assuming $\lambda_1$ is solely generated by loops involving $(\lambda_{2})_{\mu e}=(\lambda_{3})_{\mu e}$. The excluded regions are shown in gray.  Corresponding colored dashed lines display values of $\Lambda$ and $M_{\text{fermion}}$ where $\mathcal{B}r(h\rightarrow \mu e)$ is satisfied right at its experimental bound.}
\label{Fig:MuEeffective}
\end{figure}

Next, for the same choices of $(\lambda_{2,3})_{\mu e}$, we calculate the bounds in the $(\Lambda, M_{\text{fermion}})$ plane coming from the current limit on $h \to \mu e$, $\mathcal{B}r(h \rightarrow \mu e)<3.6\times 10^{-4}$~\cite{Vanhoefer:2103729}. These bounds are indicated by the dashed lines on Fig.~\ref{Fig:MuEeffective}. For $(\lambda_{2,3})_{\mu e}$ = 0.2 the bounds from $\mu \to e \gamma$ are relatively loose and, for $\Lambda > 50$ TeV, there is parameter space where the $h \to \mu e$ bound is more constraining than $\mu \to e \gamma$. However, as $(\lambda_{2,3})_{\mu e}$ is increased, we have to go to significantly higher $\Lambda$ to find the region where $h \to \mu e$ gives the stronger bound. 
\end{itemize}


\subsection{Constraints by $l_{i} \rightarrow l_{j}l_{j}l_{j}$ and nuclear $\mu-e$ conversion}
\label{subsec:MuTo3e}

We now go back to the generic flavor scenario in which none of the $(\lambda_{k})_{ij}$ is null, continuing the analysis done in Sec.~\ref{subsec:MuEGamma}. There are LFV limits on processes other than the radiative decays (generated by dipole operators) examined above, such as the $l_{i}\rightarrow l_{j}l_{j}l_{j}$ and $\tau^{-} \rightarrow l_{j}^{-}l_{k}^{+}l_{k}^{-}$ decays and the muon-electron conversion in nuclei. 

Within our setup, there are two ways of generating the amplitude $l_{i}\rightarrow l_{j}l_{j}l_{j}$: i.) we can either start with the $l_{i}\rightarrow l_{j}\gamma$ diagram in Fig.~\ref{Fig:basicDiagrams} and attach a $\ell^-\ell^+$ pair to the gauge boson ($Z$ is possible as well as a photon), or ii.) start with the left diagram of Fig.~\ref{Fig:basicDiagrams} and attach an $\ell^-\ell^+$ pair to the Higgs line. Both possibilities are shown in Fig.~\ref{Fig:LiTo3Lj}. The contributions involving a $Z$ or Higgs are highly suppressed, so it is sufficient to focus on $l_{i}\rightarrow l_{j}l_{j}l_{j}$ through a virtual photon. Notice that this means that $l_{i}\rightarrow l_{j}l_{j}l_{j}$ will also be controlled exclusively by the $(\lambda_{1})_{ij}$ four-fermion coupling.

Following the logic of the previous sections, we start with the $\tau-\mu$ sector\footnote{Recall that the set of couplings $\lambda_{i}$ in $\tau-\mu$ sector is independent from its counterpart in the $\mu-e$.}. Currently $\mathcal{B}r(\tau \rightarrow \mu \mu \mu)<2.1\times 10^{-8}$~\cite{Hayasaka:2010np}, which is comparable to the bound on $\tau \rightarrow \mu \gamma$. However, the fact that the $\tau \rightarrow \mu \mu \mu$ amplitude is obtained from the $\tau \rightarrow \mu \gamma$ diagram by inserting an extra electromagnetic vertex implies that the parametric $\Lambda, M_{\text{fermion}}$ dependence of the width of the former is the same as for the later, but $\mathcal{B}r(\tau \to \mu\mu\mu)$ will carry an extra factor of $\alpha_{em}$. As such, the constraints coming $\tau \rightarrow \mu \mu \mu$ are weaker than the bounds from $\tau \rightarrow \mu \gamma$ derived in Sec.~\ref{subsec:TauMuGamma}.

\begin{figure}[h!]
\centering
\includegraphics[scale=0.5,keepaspectratio=true]{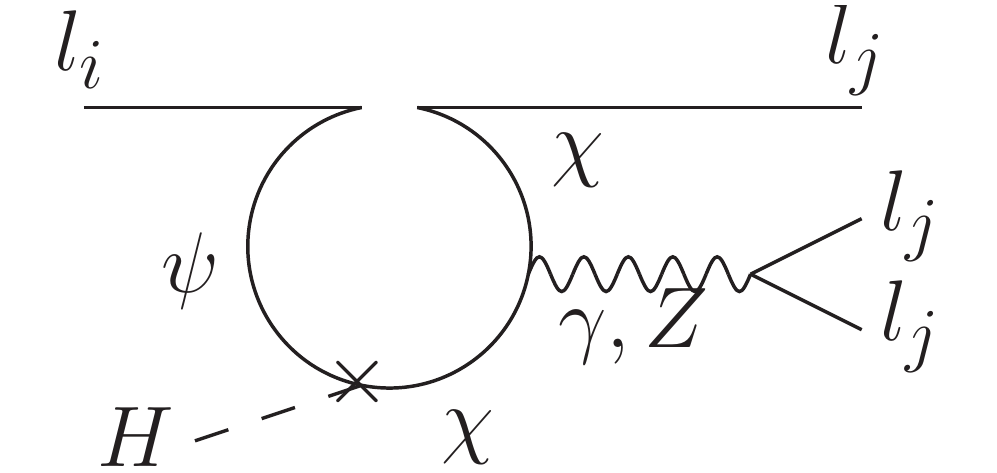}
\hspace{10mm}
\includegraphics[scale=0.5]{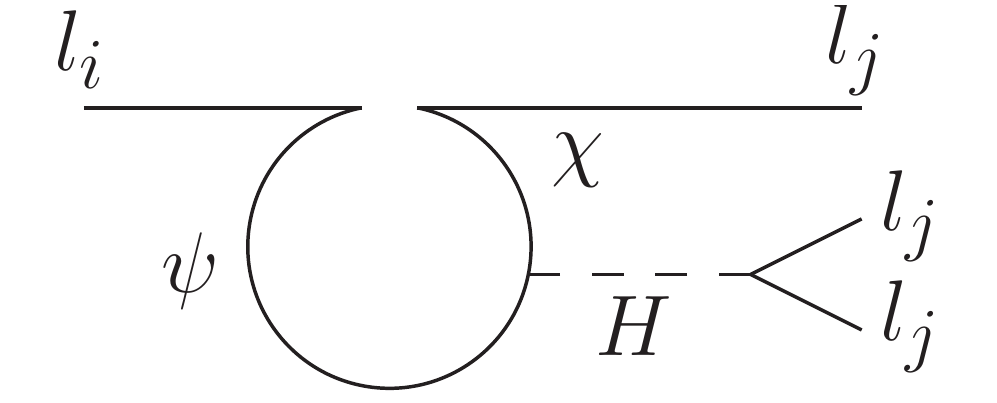}
\caption{\textbf{Left:} Leading-order amplitude for $l_{i} \rightarrow l_{j}l_{j}l_{j}$ through gauge bosons. \textbf{Right:} Amplitude $l_{i} \rightarrow l_{j}l_{j}l_{j}$ through a Higgs.}
\label{Fig:LiTo3Lj}
\end{figure}

The story in the $\mu  - e$ sector is analogous. The experimental bounds on $\mu \to eee$ are weaker than the bounds from $\mu \to e \gamma$ by an order of magnitude, $\mathcal{B}r(\mu \rightarrow eee)<1.0\times 10^{-12}$ \cite{Bertl:1985mw},  while the width for $\mu \to eee$ is suppressed by $O(\alpha_{em})$\footnote{Either $Q\alpha_{em}$ or $g_{Z}^{(\chi)}g_{Z}^{(e)}$ -- the couplings to the $Z$-- depending on which boson is attached to the loop} (again assuming the $Z$ and Higgs contributions can be neglected) compared to $\mu \to e\, \gamma$. A comparison of $\mu \rightarrow e\gamma$ with muon-electron conversion, discussed below, is shown in Fig.~\ref{Fig:MuE}.

Lastly, we examine bounds from muon-electron conversion. In muon-electron conversion, the amplitude is obtained from the exchange of either a Higgs or gauge boson ($\gamma$ or $Z$, depicted in Fig.~\ref{Fig:MuEConv}) between a loop of virtual vectorlike fermions that are connected nucleons through a four-fermion interaction. A second diagram, mediated by a Higgs, is present but can again be neglected as a consequence of the tiny up- and down-quark Yukawas. The dipole operator coefficient for our $\mu-e$ amplitude is, in the notation of \cite{Cirigliano:2009bz},
\begin{equation}
c_{\text{L(R)}}=\dfrac{2}{16\pi^{2}}\mathcal{G}(M_{\text{fermion}})\,e\,Q_{\chi}\dfrac{y(\lambda_{1})_{\mu e}v}{m_{\mu}},
\end{equation}
where $Q_{\chi}$ is $\chi$'s electric charge and $\mathcal{G}$ is the same loop function that appeared in $l_{i}\rightarrow l_{j}\gamma$. Evidently the nuclear $(\mu-e)$ conversion shares the same functional dependence on $\Lambda$ and $M_{\text{fermion}}$ as in the radiative LFV decays.

\begin{figure}[h!]
\centering
\includegraphics[scale=0.5,keepaspectratio=true]{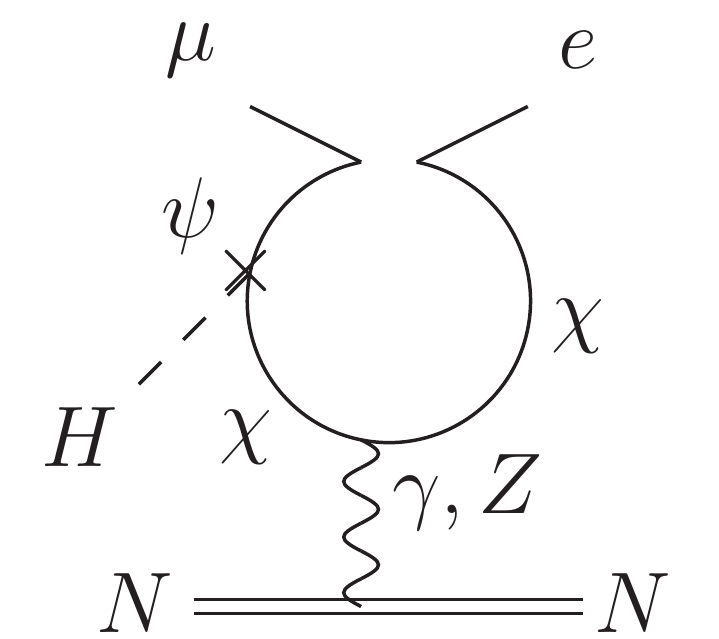}
\caption{$\mu-e$ conversion amplitude from gauge boson exchange. Analog diagrams with a Higgs connecting the loop to the nucleon are suppressed by the 1st generation quark Yukawas.}
\label{Fig:MuEConv}
\end{figure}

A detailed calculation of the rate for $\mu-e$ conversion for a general collection of LFV operators is provided in~\cite{Kitano:2002mt} and it is based on solving the Dirac equation in the external field set by the nucleus and a estimation of the relevant nuclear matrix elements. In terms of the conversion-to-capture ratio, the best current limit is $\Gamma_{\text{conversion}}^{(Z)}/\Gamma_{\text{capture}}^{(Z)}<7\times 10^{-13}~(90\%\text{ C.L.})$ in gold ($Z=79$)~\cite{Bertl:2006up}. The plot in Fig.~\ref{Fig:MuE} compares today's values for the $\mu - e$ conversion limit (thick blue dashed contour) and the $\mu \rightarrow e\gamma$ bound (red thick dashed line). The conversion bound is currently weaker than the radiative one, thus the viable parameter space is controlled by $\mu \rightarrow e\gamma$.

\begin{figure}[h!]
\centering
\includegraphics[scale=0.63,keepaspectratio=true]{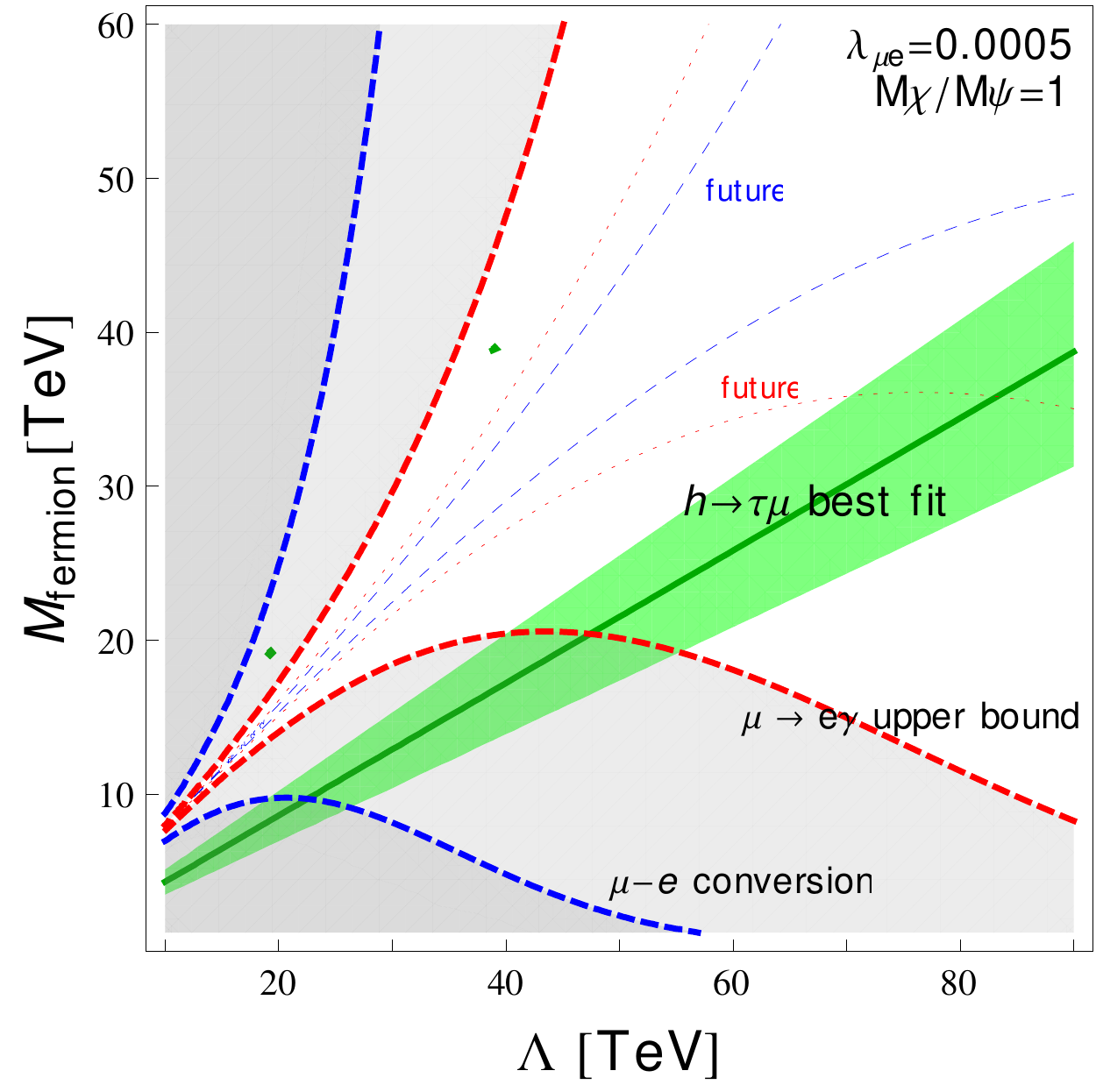}
\caption{Comparison of bounds on $\mu-e$ observables. Current limits are shown in thick dashed lines, and the thin dashed lines represent the corresponding sensitivity improvements in the future experiments of Table \ref{Tab:projections}.}
\label{Fig:MuE}
\end{figure}

\subsection{Results for future projections}
\label{subsec:Future}

New searches for the LFV processes $l_i \rightarrow l_j \gamma$, $l_i \rightarrow l_j l_k l_m$ and $\mu -e$ conversion have been planned for the near future and are expected to improve the corresponding bounds by at least one order of magnitude. For example, the MEG experiment~\cite{Adam:2013mnn} plans to improve the existing $\mathcal{B}r(\mu \rightarrow e\gamma)$ constraint to $6 \times 10^{-14}$ after a running time of 3 yr. Meanwhile, the target of Mu2e~\cite{Abrams:2012er} is sensitivity better than $10^{-15}$ in ($\mu - e$) nuclear conversion (after a similar run time). The expected future limits on the LFV processes analyzed in this work are collected in Table 1 together with their runtimes. The list -- not exhaustive -- includes representative collaborations.

\begin{table}[h!]
\begin{center}
\begin{tabular}{cccc}
\hline
\hline
Rate & Projection & Runtime [yr] & Experiment \\
\hline
    $\mu \rightarrow e\gamma$ & $\sim 6\times 10^{-14}$ & 3   & MEG~\cite{Adam:2013mnn} \\
    $\mu-e\text{ conversion}$ & $\sim 10^{-15}$         & 3   & Mu2e~\cite{Abrams:2012er} \\
$\tau \rightarrow \mu \gamma$ & $\sim 3\times 10^{-9}$  & 5   & Belle~\cite{Aushev:2010bq}\\    
        $\mu \rightarrow eee$ & $\sim 10^{-16}$         & 2.5 & Mu3e~\cite{Blondel:2013ia} \\
\hline
\hline
\end{tabular}
\caption{Expected sensitivity improvements for the LFV rates analyzed in the present model.}
\label{Tab:projections}
\end{center}
\end{table}

For $\mu \rightarrow e \gamma$ and nuclear $\mu-e$ conversion, the future bounds (in thin dashed lines) are compared to the current ones (thick dashed lines) in Fig.~\ref{Fig:MuE}. Evidently, the forthcoming limits greatly restrict the parameter space in the analyzed window, $ M_{\text{fermion}} < 50$ TeV and $20\, \text{TeV} < \Lambda < 80\, \text{TeV}$, ruling out the previous allowed ($\Lambda,M_{\text{fermion}}$) region containing the $h\rightarrow \tau\mu, y = y' = 0.25$ best fit band. In order to realign the $h\rightarrow \tau\mu$ band with the region consistent with the future bounds, we need a higher slope in the ($\Lambda,M_{\text{fermion}}$) plane, which, recalling from Section \ref{subsec:TauMuGamma}, can be accomplished by taking smaller Yukawas $y,~y'$. Of course, one may continue decreasing $\lambda_{\mu e}$ below $0.0005$ so that the $\mu \rightarrow e\gamma$ contours open up more, but this will exacerbate the hierarchy between the $\tau-\mu$ and $\mu-e$ couplings.

 If the LHC confirms the $h\rightarrow \tau\mu$ signal, we can think of two potential scenarios depending on whether or not the radiative decays $l_i\rightarrow l_j\gamma$ are found, since Higgs decays and radiative decays are governed by a different set of couplings. If either of the radiative decays $\mu\rightarrow e \gamma$ or $\tau \rightarrow \mu\gamma$ is respectively confirmed by MEG or Belle, this will imply that the corresponding coupling $\lambda_1$, whether tree-level or effective, is nonzero, as explained in Sec.~\ref{subsec:FermionmModel}. On the other hand, if $\tau \to \mu \gamma$ is not found, searches for $\mu-e$ conversion and $\mu\rightarrow eee$ at the Mu2e and Mu3e experiments are still well motivated -- as these are controlled by different couplings. Likewise, looking for $h\rightarrow \mu e$ is well motivated under our setup because, departing from conventional approaches, it is not true that the non-observation of $\mu \rightarrow e\gamma$ entails that of $h\rightarrow \mu e$ (Sec.~\ref{sec:EffectsLambda1Eff}).
 
\section{Discussion and conclusions}                 
\label{sec:Conclusion}

We have discussed two models that could explain the excess in the lepton flavor violating Higgs decay $h \rightarrow \tau \mu$ reported by ATLAS and CMS while it awaits for confirmation as a real signal or is disproved by better statistics.

Within the MSSM, LFV driven exclusively by leptonic $A$-terms is not affected by lepton Yukawa suppression, and if the wino, Higgsino, and squarks are much heavier than the bino, this is the single dominant source of LFV. In a non-exhaustive approach that ignores field renormalization diagrams, we find the relevant parameter space that can accommodate the excess while respecting the bounds for the branching ratio of $\tau \rightarrow \mu \gamma$. This is possible because one can in principle fix the branching ratio of the process $h\rightarrow \tau \mu$ using the ratio $|A_{t}| /m_{\tilde{l}}$, then vary the slepton mass $m_{\tilde{l}}$ until it satisfies the $\tau \rightarrow \mu \gamma$ bounds. However, apart from making sure other LFV observables such as $l_{i}\rightarrow l_{j}\gamma$ are respected, one must avoid configurations with large, non-diagonal $A$-terms that trigger the appearance of color- and charge-breaking minima in the scalar potential. We have shown here that, in this simple MSSM setup, LFV $A$-terms by themselves cannot accommodate the Higgs LFV excess and simultaneously respect the analytical stability bounds obtained in the literature. Therefore, if $h\rightarrow \tau \mu$ is confirmed as a signal, this decay would require a more elaborate extension (i.e. additional interactions of superfields) of the MSSM.

On the other hand, we studied a nonsupersymmetric model where an effective LFV Yukawa interaction of the Higgs is formed through a loop of vectorlike fermions attached to the SM leptons by a four-fermion interactions. This setup has been shown to be capable of fitting the $h\rightarrow \tau \mu$ excess while satisfying the bounds from radiative leptonic LFV decays in the $\tau-\mu$ and $\mu-e$ sectors under a simplified set of assumptions: (i) that we make an educated choice of the hypercharge of the new states (avoiding mixing between the new fermions and the SM leptons, and keeping the number of new operators as low as possible), and (ii) we do not favor specific UV realizations that could set any of the four-fermion couplings to zero. Our setup satisfies the mentioned bounds provided that the high scale $\Lambda$ starts from 35 TeV and the heavy fermion masses are (quasi)degenerate and above 15 TeV (for Yukawa couplings $y = y' \sim 0.25$). Probing these fermions directly is beyond the current reach of the LHC.

 Moreover, since the radiative LFV decays and Higgs decays are controlled by different couplings, the fact that our model reproduces $h\rightarrow \tau \mu$ does not necessarily imply a definite prediction on $l_{i}\rightarrow l_{j}\gamma$, though this decay can be induced at loop level even if the coupling $(\lambda_1)_{\tau\mu} = 0$ at tree level. Therefore, even if the LHC confirms the excess in $h\rightarrow\tau\mu$, other lepton violation observables will be needed to determine the different parameters of our effective Lagrangian and the UV physics behind it.

In conclusion, our investigation has shown that any model utilizing extra scalars to explain the lepton flavor violation Higgs decays will run into problems of stability due to the size of the couplings needed, and that models with vectorlike fermions will have to be of the kind studied in this paper, i.e., ones that can only mix with SM fermions through higher dimensional operators to avoid dangerous contributions to flavor changing neutral currents.

\acknowledgments
This research was supported in part by the National Science Foundation under Grants
PHY-1417118 and PHY-1520966. We would like to thank Wolfgang Altmannshofer, Stefania Gori, Alex Kagan, Jure Zupan,  Daniel Aloni, Yosef Nir, Emmanuel Stamou, Ernesto Arganda, Maria Jos\'e Herrero, Reynaldo Morales, and Alejandro Szynkman for relevant discussions.

\appendix                                                      
\section{Appendix: Loop functions}
\label{sec:AppendixFunctions}

The loop function appearing in (\ref{WidthAtermhTauMu}) in the $A$-term model is given by
\begin{equation}
H(r)=\dfrac{-1+r^{2}-\log{(r^2)}}{(-1 + r^{2})^{2}},
\end{equation}
with $r\equiv m_{\widetilde{l}}/m_{\widetilde{B}}$. In the four-fermion interaction model the loop functions are defined by
\begin{align}
\mathcal{H}_{\text{closed}}(M_{\psi},M_{\chi})
&= \dfrac{-2}{M_{\chi}^{2}(M_{\chi}^{2}-M_{\psi}^{2})}\biggl\{ M_{\chi}^4-M_{\psi}^4+\biggl[M_{\chi}^4\log{\left( \dfrac{\Lambda^2}{M_{\chi}^2} \right)}-M_{\psi}^4\log{\left( \dfrac{\Lambda^2}{M_{\psi}^2} \right)}\biggr] \biggr\}, \\ & \notag \\
\mathcal{G}(M_{\psi},M_{\chi})
&= \dfrac{-1}{36(M_{\chi}^{2}-M_{\psi}^{2})^{3}}\biggl[ M_{\chi}^{6}+18M_{\chi}^{4}M_{\psi}^{2}-45M_{\chi}^{2}M_{\psi}^{4}+26M_{\psi}^{6} \biggr. \notag \\
&+ \biggl. 6(2M_{\chi}^{3}-3M_{\chi}M_{\psi}^{2})^{2}\log{\left( \dfrac{M_{\psi}^{2}}{M_{\chi}^{2}} \right)}-24(M_{\chi}^{2}-M_{\psi}^{2})^{3}\log{\left( \dfrac{\Lambda^{2}}{M_{\psi}^{2}} \right)} \biggr] \label{Gfunction}.
\end{align}

The loop function $\mathcal{H}_{\text{open}}(M_{\psi},M_{\chi})$ of the open-line diagram has a dominant part which goes as $-(1/2)\mathcal{H}_{\text{closed}}(M_{\psi},M_{\chi})$ and a subdominant one proportional to the external momentum of the Higgs.  As mentioned in Section \ref{subsec:FermionmModel} the rate $\Gamma(h\rightarrow l_{i}l_{j})$ receives contributions from the closed- and open-fermion line diagrams, respectively parametrized by $\mathcal{H}_{\text{closed}}$ and $\mathcal{H}_{\text{open}}$.

\section{Appendix: Hypercharge choice}
\label{sec:AppendixChoice}

In principle for the vectorlike fermions $\chi=(\bold{1},x)$ and $\psi=(\bold{2},1/2+x)$ the hypercharge assignment $x$ can take any value. However, several issues demand a careful choice of $x$:

\textit{1. No Majorana fermions.} The hypercharge choice $x=0$ is not considered in order to avoid dealing with a Majorana fermion $\chi$ singlet of the SM gauge group. In such a case Majorana masses would be allowed in addition to the vectorlike ones.

\textit{2. Number of effective operators.} Given the hypercharges of the SM fields, the choice $x=-1$ generates five effective four-fermion operators\footnote{The parentheses refer to spinor contractions and proper $SU(2)_L$ contractions are implicit.}
\begin{equation}
(\psi e^{cj})(\chi^{c\dag}L^{i\dag}),\,\, (\psi^c L_i)(\chi^{\dag}e_j^{c\dag}),\,\,
(\psi \chi^{c})(L^{i\dag}e_j^{c\dag}),\,\, (\psi^c L_i)(\chi e^{cj}),\,\,
(\psi^c \chi)(L_i e^{cj}),
\end{equation}
whereas for $x=2$, only the last three operators above are generated.

\textit{3. Heavy charged particle.} Since $\chi$ is a $SU(2)_L$ singlet, its electric charge is directly set by the value of $x$. Any non-integer value for $x$ would allow for a stable, electrically charged particle. Given that $\chi$ can be $\mathcal{O}(15\text{ TeV})$, noninteger $x$ is ruled out due to cosmological considerations on the barionic density.

\textit{4. Chiral-vector mixing.} In the case $x=1$, the fermions $\chi$ and $\psi$ look like a fourth family of leptons. This implies that the following operators are allowed

\begin{equation}
z H^{\dag}L_{i}\chi, \,\,\, z' H\psi^{c}e^{c}, \,\,\, \bold{m}_{ij} \chi^{ci}e^{cj},
\label{mixtures}
\end{equation}

where $z$ and $z'$ are new Yukawa couplings and $\bold{m}_{ij}$ is a mixing matrix. These operators induce mass mixing between the new fermions $\chi$, $\psi$ and the SM fermions directly and upon electroweak symmetry breaking, which modifies the SM lepton masses. In order to avoid deviating from the tau mass established value, this mixing is suppressed by enforcing $zv,z'v\ll M_{\psi,\chi}$ (i.e. $z$ and $z'$ can by bounded for a given $M_{\psi,\chi}$). However, for masses $M_{\psi,\chi}$ at the $\mathcal{O}(\text{TeV})$ scale, the $\mathcal{B}r(h\rightarrow \tau \mu$) is already suppressed and the best fit (\ref{BFcombined}) cannot be achieved. A simple way to avoid these effects and an unnecessary enlargement of the parameter space, is to choose $x\neq1$ such that the operators (\ref{mixtures}) are not generated. In this case, all LFV effects are a result of the four-fermion interaction.

\bibliographystyle{JHEP}
\bibliography{References}

\end{document}